\documentclass[twocolumn]{svjour3}
\usepackage{graphicx}
\usepackage{subfigure,graphicx}
\usepackage{amsmath}
\usepackage[]{placeins}
\usepackage{amsfonts}
\usepackage{float}
\usepackage{amssymb}
\usepackage{mathtools}
\usepackage{tensor}
\usepackage{enumitem}

\def\bfu{{\bf u}}

\def\bfx{{\bf x}}

\def\bfE{{\bf E}}

\def\bfI{{\bf I}}

\def\bfN{{\bf N}}

\def\bfX{{\bf X}}

\def\e0{\varepsilon_0}

\def\s0{\sigma_0}

\long\def\symbolfootnote[#1]#2{\begingroup%
\def\thefootnote{\fnsymbol{footnote}}\footnote[#1]{#2}\endgroup}

\newcommand\sts{\sigma_{\texttt{ts}}}
\newcommand\scs{\sigma_{\texttt{cs}}}

\long\def\symbolfootnote[#1]#2{\begingroup%
\def\thefootnote{\fnsymbol{footnote}}\footnote[#1]{#2}\endgroup}

\begin{document}

\titlerunning{The revisited phase-field approach to brittle fracture: Application to indentation and notch problems}

\title{The revisited phase-field approach to brittle fracture: Application to indentation and notch problems}

\author{A. Kumar \and K. Ravi-Chandar \and   O. Lopez-Pamies}

\institute{
           Aditya Kumar \at Department of Civil and Environmental Engineering, University of Illinois, Urbana--Champaign, IL 61801-2352, USA\\
           \email{akumar51@illinois.edu}\vspace{0.1cm}
           \and
           K. Ravi-Chandar \at
           Department of Aerospace and Engineering Mechanics, The University of Texas at Austin, TX  78712--1221, USA \\
           \email{ravi@utexas.edu}
           \and
           Oscar Lopez-Pamies \at
           Department of Civil and Environmental Engineering, University of Illinois, Urbana--Champaign, IL 61801-2352, USA  \\
           D\'epartement de M\'ecanique, \'Ecole Polytechnique, 91128 Palaiseau, France \\
           \email{pamies@illinois.edu}
           }

\maketitle

\begin{abstract}

In a recent contribution, Kumar, Bourdin, Francfort, and Lopez-Pamies (J. Mech. Phys. Solids 142:104027, 2020) have introduced a comprehensive macroscopic phase-field theory for the nucleation and propagation of fracture in linear elastic brittle materials under arbitrary quasistatic loading conditions. The theory can be viewed as a natural generalization of the phase-field approximation of the variational theory of brittle fracture of Francfort and Marigo (J. Mech. Phys. Solids 46:1319--1342, 1998) to account  for the material strength at large. This is accomplished by the addition of an external driving force --- which physically represents the macroscopic manifestation of the presence of inherent microscopic defects in the material --- in the equation governing the evolution of the phase field. The main purpose of this paper is to continue providing validation results for the theory by confronting its predictions with direct measurements from three representative types of experimentally common yet technically challenging problems: $i$) the indentation of glass plates with flat-ended cylindrical indenters and the three-point bending of $ii$) U-notched and $iii$) V-notched PMMA beams.


\end{abstract}

\vspace{-0.5cm}

\section{Introduction} \label{Introduction}

In the first part of a recent contribution, Kumar et al. (2020) have argued that any formulation that aims at providing a complete macroscopic theory of nucleation and propagation of fracture in \emph{homogeneous} elastic brittle materials must account for three material inputs:
\begin{enumerate}

\item{the elasticity of the material,}

\item{its strength at large, and}

\item{its critical energy release rate.}

\end{enumerate}
This is because --- based on the myriad of experimental observations that have been amassed over the past hundred years on numerous nominally brittle ceramics, metals, and polymers alike --- Kumar et al. (2020) posit that \emph{nucleation of fracture}
\renewcommand{\labelitemi}{$\bullet$}
\begin{itemize}

\item{in the bulk is governed by the strength of the material,}

\item{from large\footnote{``Large'' refers to large relative to the characteristic size of the underlying heterogeneities in the material under investigation. By the same token, ``small'' refers to sizes that are of the same order or just moderately larger than the sizes of the heterogeneities.} pre-existing cracks is governed by the Griffith competition between the elastic energy and fracture energy,}

\item{from boundary points, be them smooth or sharp, and small pre-existing cracks is governed by the interaction among the strength, the elastic energy, and the fracture energy of the material,}

\end{itemize}
while \emph{propagation of fracture}
\begin{itemize}

\item{is, akin to nucleation from large pre-existing cracks, also governed by the Griffith competition between the elastic and fracture energies.}

\end{itemize}
The term homogeneous merits explicit clarification. It refers, in the usual manner, to materials for which the above three inputs $i$ through $iii$ can be considered as \emph{intrinsic properties} and thus independent of the geometry of the structural problem at hand. This requires that the underlying heterogeneities in the material, including its inherent defects from which fracture may originate, must be ``much smaller'' than the characteristic length scale of the structure and the scale of variation of the applied loads. At present, the precise meaning of the qualifier ``much smaller'' has been well established only for the first of the above three material inputs, and this just for linear elastic materials. For those, direct calculations show that the elasticity of heterogeneous materials wherein the length scale of the underlying heterogeneities is a mere 4 times smaller than the structural size can already be treated effectively as homogeneous; see, e.g., Drugan and Willis (1996). The situation for the strength and the critical energy release rate is more delicate and not yet settled. Numerous efforts have been and continue to be devoted to gaining insight into the latter; see, e.g., the works of Gao and Rice (1989), Bower and Ortiz (1991), Cox and Yang (2006), Hossain et al. (2014), and Hsueh and Bhattacharya (2016). On the other hand, much less work has been dedicated to pinpointing when the strength of a material may be considered as an intrinsic property, presumably because of the technical difficulties of carrying out experiments where the sizes of the inherent defects --- both within the bulk and on the boundary of any given piece of material --- are controlled with sufficient accuracy, and, on the theoretical front, also because of the lack of appropriate mathematical definitions of defects and of their homogenization.

In the most basic setting, that of homogeneous isotropic linear elastic brittle materials, the argument of Kumar et al. (2020) entails precisely that any formulation that aims at providing a complete macroscopic theory of nucleation and propagation of fracture must incorporate as material inputs the stored-energy function
\begin{equation}
W(\bfE)=\mu \, {\rm tr}\,\bfE^2+\dfrac{\lambda}{2}\left({\rm tr}\,\bfE\right)^2\label{Wiso}
\end{equation}
describing the elastic response of the material, the strength surface
\begin{equation}
\mathcal{F}(\sigma_1,\sigma_2,\sigma_3)=0 \label{Fiso}
\end{equation}
describing its strength under arbitrary uniform stress conditions, and the critical energy release rate
\begin{equation}
G_c \label{Gciso}
\end{equation}
describing the growth of cracks within it. Making use of standard notation, the coefficients $\lambda$ and $\mu$  in expression (\ref{Wiso}) stand for the first and second Lam\'e moduli of the material. In expression (\ref{Fiso}), $\sigma_1$, $\sigma_2$, $\sigma_3$ stand for the eigenvalues of the Cauchy stress tensor $\boldsymbol{\sigma}$, that is, the principal Cauchy stresses. The definition of the strength surface (\ref{Fiso}) in terms of these arguments is as follows. When any piece of the isotropic linear elastic material of interest is subjected to a state of monotonically increasing \emph{uniform} but otherwise arbitrary stress, fracture will nucleate from one or more of its inherent defects at a critical value of the applied stress. The set of all such critical stresses defines a surface in stress space. In terms of the Cauchy stress tensor $\boldsymbol{\sigma}$, that surface is given by (\ref{Fiso}). Because of the assumption of material isotropy invoked here, the dependence on $\boldsymbol{\sigma}$ enters only via $\sigma_1$, $\sigma_2$, $\sigma_3$.

At this point, it is important to emphasize that standardized tests to measure the elastic material constants $\lambda$ and $\mu$ and the critical energy release rate $G_c$ for a given material of interest have long been available and can now be readily carried out with conventional equipment; see, e.g., Chapter 6 in the monograph by Zehnder (2012). By contrast, it is extremely difficult to carry out experiments that probe the entire space of uniform stresses in order to measure the entire strength surface (\ref{Fiso}) for a given material of interest. Indeed, most of the experimental strength data available in the literature is narrowly restricted to uniaxial tensile and compressive strength when the state of uniform stress is of the form $\boldsymbol{\sigma}={\rm diag}(\sigma_{1},0,0)$ with $\sigma_{1}>0$ and $\sigma_{1}<0$. To a lesser extent, there is also strength data available for the more general subset of plane-stress conditions when $\boldsymbol{\sigma}={\rm diag}(\sigma_{1},\sigma_{2},0)$, which includes as a special case shear strength data when $\sigma_{2}=-\sigma_{1}$; see, e.g., Chapter 10 in the book by Munz and Fett (1999) and references therein. Another key difference between $\lambda$, $\mu$, $G_c$, and the strength surface (\ref{Fiso}) is that the latter is inherently stochastic. This is because the strength at a macroscopic material point depends on the nature of the underlying defects from which fracture initiates, and this is known to exhibit a stochastic spatial variation in any given piece of material. This spatial variation is most acute when comparing material points within the bulk of the given piece with material points on its boundary, since different fabrication processes or boundary treatments (such as polishing or chemical treatments) can drastically affect the nature of boundary defects vis-\`a-vis those in the bulk.

In the latter part of their contribution, having pinpointed the above-enumerated requirements, Kumar et al. (2020) introduced a comprehensive macroscopic theory of nucleation and propagation of fracture --- regularized, of phase-field type --- in linear elastic brittle materials under arbitrary quasistatic loading conditions that incorporates directly (\ref{Wiso}), (\ref{Fiso}), (\ref{Gciso}) as the material inputs. The theory corresponds to a generalization of the phase-field regularization (Bourdin et al., 2000) of the variational theory of brittle fracture of Francfort and Marigo (1998), which in turn corresponds to the mathematical statement of Griffith's fracture postulate in its general form of energy cost-benefit analysis (Griffith, 1921). In the footstep of Kumar et al. (2018a), the generalization amounts: $i)$ to considering the Euler-Lagrange equations of the standard phase-field regularization --- and \emph{not} the variational principle itself --- as the primal model and $ii)$ to adding an external driving force in the Euler-Lagrange equation governing the evolution of the phase field to describe the macroscopic manifestation of the presence of the inherent defects in the material, that is, its strength at large.

As a first validation step, Kumar et al. (2020) also provided in their paper direct comparisons between predictions obtained from the theory and experimental results on a wide range of materials (titania, graphite, polyurethane, PMMA, and alumina) under loading conditions that spanned the full range of fracture nucleation settings (within the bulk, from large pre-existing cracks, boundary points, and small pre-existing cracks).

The main purpose of this paper is to continue providing validation results for the theory of Kumar et al. (2020). We do so by confronting its predictions with representative experimental results for the nucleation and propagation of fracture in indentation and notch problems. We focus in particular on the indentation of glass plates with flat-ended cylindrical indenters and the three-point bending of U-notched and V-notched PMMA beams. The rationale for our choice is twofold. On the one hand, these types of problems are very common in the experimental literature. On the other hand, their analysis is technically challenging because of the singular or high-gradient elastic fields that they feature prior to the nucleation of fracture. We begin in Sections \ref{Sec: Approach} and \ref{Sec: Driving Force} by summarizing the general fracture theory of Kumar et al. (2020) and its specialization to the prototypical case of Drucker-Prager strength surfaces. We then present its application in Sections \ref{Sec: Indentation}, \ref{Sec: U-Notch}, and \ref{Sec: V-Notch} to the indentation, U-notch, and V-notch problems, respectively. We close by recording some concluding remarks in Section \ref{Sec: Final Comments}.

\section{The revisited phase-field approach to brittle fracture} \label{Sec: Approach}

Consider a structure made of an isotropic linear elastic brittle material, with stored-energy function (\ref{Wiso}), strength surface (\ref{Fiso}), and critical energy release rate (\ref{Gciso}), that occupies an open bounded domain $\mathrm{\Omega}\subset \mathbb{R}^3$, with boundary $\partial\mathrm{\Omega}$ and unit outward normal $\bfN$, in its undeformed and stress-free configuration at time $t=0$. At a later time $t \in (0, T]$, due to an externally applied displacement $\overline{\bfu}(\bfX, t)$ on a part $\partial\mathrm{\Omega}_\mathcal{D}$ of the boundary and a traction $\overline{\textbf{t}}(\bfX,t)$ on the complementary part $\partial\mathrm{\Omega}_\mathcal{N}=\partial\mathrm{\Omega}\setminus \partial\mathrm{\Omega}_\mathcal{D}$, the position vector $\bfX$ of a material point moves to a new position specified by
\begin{equation*}
\bfx=\bfX+\bfu(\bfX,t)
\end{equation*}
in terms of the displacement field $\bfu$. We write the infinitesimal strain tensor as
\begin{equation*}
\bfE(\bfu)=\dfrac{1}{2}(\nabla\bfu+\nabla\bfu^T).
\end{equation*}
In response to the same externally applied mechanical stimuli that result in the above-described deformation, cracks can also nucleate and propagate in the structure. Those are described in a regularized fashion by the order parameter or phase field
\begin{equation*}
v=v(\bfX,t)
\end{equation*}
taking values in $[0,1]$. Precisely, $v=1$ identifies regions of the sound material, whereas $v=0$ identifies regions of the material that have been fractured.

According to the theory of Kumar et al. (2020), the displacement field $\bfu_k(\bfX)=\bfu(\bfX,t_k)$ and phase field $v_k(\bfX)=v(\bfX,t_k)$ at any material point $\bfX\in\overline{\mathrm{\Omega}}$ and discrete time $t_k\in\{0=t_0,t_1,...,t_m,$ $t_{m+1},...,$ $t_M=T\}$ are determined by the system of coupled partial differential equations (PDEs)
\begin{equation}
\left\{\begin{array}{ll}
 {\rm Div}\left[v_{k}^2\dfrac{\partial W}{\partial \bfE}(\bfE(\bfu_{k}))\right]={\bf0},\quad \bfX\in\mathrm{\Omega},\\[10pt]
\bfu_{k}=\overline{\bfu}(\bfX,t_{k}),\quad \bfX\in\partial  \mathrm{\Omega}_{\mathcal{D}},\\[10pt]
 \left[v_{k}^2\dfrac{\partial W}{\partial \bfE}(\bfE(\bfu_{k}))\right]\bfN=\overline{\textbf{t}}(\bfX,t_{k}),\quad \bfX\in\partial \mathrm{\Omega}_{\mathcal{N}}\end{array}\right. \label{BVP-u-theory}
\end{equation}
and
\begin{equation}
\left\{\begin{array}{l}
\varepsilon \, G_c \triangle v_{k}=\dfrac{8}{3}v_{k} W(\bfE(\bfu_{k}))-\dfrac{4}{3}c_\texttt{e}(\bfX,t_{k})-\dfrac{G_c}{2\varepsilon},
\\[5pt] \mbox{if } v_{k}(\bfX)< v_{k-1}(\bfX),\quad \bfX\in\mathrm{\Omega} \\[10pt]
\varepsilon \, G_c \triangle v_{k}\geq\dfrac{8}{3}v_{k} W(\bfE(\bfu_{k}))-\dfrac{4}{3}c_\texttt{e}(\bfX,t_{k})-\dfrac{G_c}{2\varepsilon},
\\[5pt] \mbox{if } v_{k}(\bfX)=1\; \mbox{ or }\; v_{k}(\bfX)= v_{k-1}(\bfX)>0,\quad \bfX\in\mathrm{\Omega} \\[10pt]
v_{k}(\bfX)=0,\quad \mbox{ if } v_{k-1}(\bfX)=0,\quad \bfX\in\mathrm{\Omega}
\\[10pt]
\nabla v_{k}\cdot\bfN=0,\quad \bfX\in \partial\mathrm{\Omega}
   \end{array}\right. \label{BVP-v-theory}
\end{equation}
with $\bfu(\bfX,0)\equiv\textbf{0}$ and $v(\bfX,0)\equiv1$, where $\nabla\bfu_k(\bfX)=\nabla\bfu(\bfX,t_k)$, $\nabla v_k(\bfX)=\nabla v(\bfX,t_k)$, $\triangle v_k(\bfX)= \triangle v(\bfX,$ $t_k)$, and where $\varepsilon>0$ is a regularization or localization length and $c_\texttt{e}(\bfX,t)$ is a driving force whose specific constitutive prescription depends on the particular form of the strength surface (\ref{Fiso}). In the next section, we spell out a specific form for $c_\texttt{e}(\bfX,t)$ for the prototypical case of Drucker-Prager strength surfaces.

\begin{remark}{\rm The localization length $\varepsilon$ in equations (\ref{BVP-u-theory})--(\ref{BVP-v-theory}) is just a regularization parameter that is void of any further physical meaning. In practice, it should be selected to be smaller than the smallest characteristic length scale in the structural problem at hand.
}
\end{remark}

\begin{remark}{\rm The inequalities in (\ref{BVP-v-theory}) embody the classical assumption also adopted here that fracture is a purely dissipative and irreversible
process. As elaborated in Section 2 of Kumar et al. (2018a), however, a generalization that would account for the possibility of healing is straightforward; see also Francfort et al. (2019).
}
\end{remark}

\begin{remark}{\rm On their own, equations (\ref{BVP-u-theory}) and (\ref{BVP-v-theory}) are second-order elliptic PDEs for the displacement field $\bfu$ and the phase field $v$. Accordingly, their numerical solution is amenable to a standard finite-element staggered scheme in which (\ref{BVP-u-theory}) and (\ref{BVP-v-theory}) are discretized with finite elements and solved iteratively one after the other at every time step $t_k$ until convergence is reached. All the simulations presented in this paper are generated with such a scheme.
}
\end{remark}

\section{The external driving force $c_{\texttt{e}}$ for Drucker-Prager strength surfaces} \label{Sec: Driving Force}

In practice, as alluded to above, the strength surface (\ref{Fiso}) of a given material of interest is only partly known. In fact, often times, only its uniaxial tensile and compressive strengths are freely available in the literature. To deal with this lack of experimental results, one usually resorts to the use of a model that can fit and extrapolate the available strength data to the entire stress space. In all the simulations that we present below, among a plurality of possibilities --- see, e.g., Rankine (1857), Drucker and Prager (1952), Bresler and Pister (1958), Willam and Warnke (1975), and Christensen et al. (2002) to name a few --- we make use of Drucker-Prager strength surfaces:
\begin{equation}
\mathcal{F}=\sqrt{J_2}+\gamma_1 I_1+\gamma_0=0.\label{F-DP}
\end{equation}
In this expression,
\begin{align}
&I_1={\rm tr}\,\boldsymbol{\sigma}=\sigma_1+\sigma_2+\sigma_3,\nonumber\\
&J_2=\dfrac{1}{2}{\rm tr}\,\boldsymbol{\sigma}^2_D=\dfrac{1}{6}\left((\sigma_1-\sigma_2)^2+(\sigma_1-\sigma_3)^2+\right.\nonumber\\
&\hspace{2.3cm}\left.(\sigma_2-\sigma_3)^2\right), \label{T-invariants}
\end{align}
where $\boldsymbol{\sigma}_D=\boldsymbol{\sigma}-1/3\left({\rm tr}\,\boldsymbol{\sigma}\right)\bfI$, and $\gamma_0$ and $\gamma_1$ are two material constants. In the sequel, we shall calibrate these constants with the uniaxial tensile and compressive strengths, say $\sts$ and $\scs$, of the given material. They then specialize to
\begin{align*}
\gamma_0=-
\dfrac{2\scs\sts}{\sqrt{3}(\sigma_{\texttt{cs}}+\sigma_{\texttt{ts}})}
\end{align*}
and
\begin{align*}
\gamma_1=\dfrac{\sigma_{\texttt{cs}}-\sigma_{\texttt{ts}}}{\sqrt{3}(\sigma_{\texttt{cs}}+\sigma_{\texttt{ts}})}.
\end{align*}

\begin{remark}{\rm The two-material-parameter strength surface (\ref{F-DP}), originally introduced by Drucker and Prager (1952) to model the yielding of soils, is arguably the simplest model that is capable of describing reasonably well the strength of many nominally brittle materials; see Chapter 10 in Munz and Fett (1999) and Section 2 in Kumar et al. (2020), for instance.
}
\end{remark}

Having settled on the choice of Drucker-Prager strength surfaces (\ref{F-DP}), we follow the blueprint provided by Kumar et al. (2020) for constructing external driving forces $c_{\texttt{e}}$ and set
\begin{align}
c_{\texttt{e}}(\bfX,t)=&\widehat{c}_{\texttt{e}}(I_1,J_2;\varepsilon)\nonumber\\
=&\beta_2^\varepsilon\sqrt{J_2}+\beta_1^\varepsilon I_1+\beta_0^\varepsilon+\nonumber\\
&\left(1-\dfrac{\sqrt{I_1^2}}{I_1}\right)\left(\dfrac{J_2}{2\mu}+\dfrac{I_1^2}{6(3\lambda+2\mu)}\right),\label{cehat}
\end{align}
where
\begin{equation}
\left\{\begin{array}{l}\beta_0^\varepsilon=\delta^\varepsilon\dfrac{3 G_c}{8 \varepsilon}\\[12pt]
\beta^\varepsilon_1=-\left(\dfrac{(1+\delta^\varepsilon)(\scs-\sts)}{2\scs\sts}\right)\dfrac{3G_c}{8\varepsilon}+\nonumber\\[8pt]
\hspace{1cm}\dfrac{\sts}{6(3\lambda+2\mu)}+\dfrac{\sts}{6\mu}\\[12pt]
\beta^\varepsilon_2=-\left(\dfrac{\sqrt{3}(1+\delta^\varepsilon)(\scs+\sts)}{2\scs\sts}\right)\dfrac{3G_c}{8\varepsilon}+\nonumber\\[8pt]
\hspace{1cm}\dfrac{\sts}{2\sqrt{3}(3\lambda+2\mu)}+
\dfrac{\sts}{2\sqrt{3}\mu}\end{array}\right., \label{betas}
\end{equation}
$I_1$ and $J_2$ stand for the invariants (\ref{T-invariants}) of the Cauchy stress
\begin{equation*}
\boldsymbol{\sigma}(\bfX,t)=v^2\dfrac{\partial W}{\partial \bfE}(\bfE(\bfu))
\end{equation*}
and, hence, read as
\begin{equation*}
I_1=(3\lambda+2\mu) v^2 {\rm tr}\,\bfE(\bfu)\quad {\rm and}\quad J_2=2\mu^2 v^4 {\rm tr}\,\bfE^2_D(\bfu)
\end{equation*}
with $\bfE_D(\bfu)=\bfE(\bfu)-1/3\left({\rm tr}\,\bfE(\bfu)\right)\bfI$ in terms of the displacement field $\bfu$ and phase field $v$, and where $\delta^\varepsilon$ is a unitless $\varepsilon$-dependent coefficient whose calibration needs to be carried out numerically. Precisely, as elaborated in Subsection 4.3.2 in Kumar et al. (2020), for a given set of material constants $\lambda$, $\mu$, $G_c$, $\sigma_{\texttt{ts}}$, $\sigma_{\texttt{cs}}$, and a given finite localization length $\varepsilon$, the value of $\delta^\varepsilon$ is determined by considering any boundary-value problem of choice for which the nucleation from a large pre-existing crack can be determined exactly --- according to Griffith's sharp theory of brittle fracture for linear elastic materials (LEFM) --- and then by having the phase-field theory (\ref{BVP-u-theory})--(\ref{BVP-v-theory}) with external driving force (\ref{cehat}) match that exact solution thereby determining $\delta^\varepsilon$.

\begin{remark}{\rm As required by the construction process laid out by Kumar et al. (2020), the external driving force (\ref{cehat}) is asymptotically identical in the limit as $\varepsilon\searrow 0$ to that utilized by Kumar et al. (2020) in their comparisons with experiments, but differs from it in that it has an additional correction of $O(\varepsilon^0)$: its last term $(1-\sqrt{I_1^2}/I_1)(J_2/2\mu+I_1^2/6(3\lambda+2\mu))$. This additional correction is non-zero only when $I_1<0$ and we include it here in order to have an improved description of the compressive part (when $I_1<0$) of the given material strength surface (\ref{F-DP}) for a larger range of finite values of the localization length $\varepsilon$.
}
\end{remark}

\section{Application to the indentation of glass plates with flat-ended cylindrical indenters} \label{Sec: Indentation}

\begin{figure}[t!]
		\centering
		\includegraphics[width=0.85\linewidth]{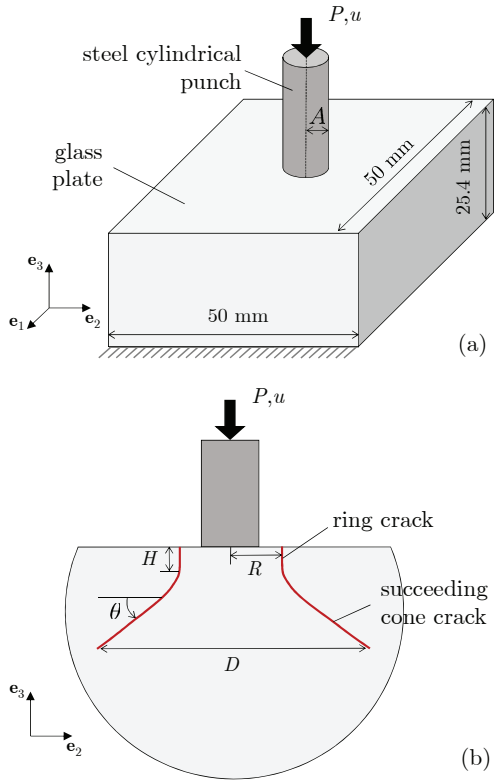}\vspace{-0.1cm}
		\caption{Schematics (a) of the initial specimen geometry and applied boundary conditions for the indentation experiments carried out by Mouginot and Maugis (1985) on borosilicate glass and (b) of the geometry of the cracks that they observed at large enough applied displacements $u$.}\label{Fig1}
\end{figure}

In this section, we confront the predictions generated by the fracture phase-field theory (\ref{BVP-u-theory})--(\ref{BVP-v-theory}) with external driving force (\ref{cehat}) to the experiments of Mouginot and Maugis (1985) for the indentation of borosilicate glass plates with flat-ended steel cylindrical indenters. As schematically depicted in Fig. \ref{Fig1}(a), these authors indented $50 \,{\rm mm}\, \times 50 \,{\rm mm}\, \times 25.4 \,{\rm mm}$ glass plates with flat punches of six different radii, $A=0.05, 0.10, 0.25, 0.50, 1.00,$ and $2.50$ mm. The specimens were indented at the fixed displacement rate of $\dot{u}=0.83$ $\mu$m/s. Prior to their indentation, the surfaces of the specimens were abraded with different grades of abrasive paper or diamond paste. The most complete set of results that were reported, and hence the one for which we carry out the comparisons here, pertains to specimens that were abraded with a 1000 grit silicon carbide paper (labeled SiC 1000).

Before proceeding with the comparisons \emph{per se}, it is pertinent to mention that since the pioneering experiments of Hertz (1882) a multitude of investigators have studied how fracture nucleates and propagates in many nominally brittle ceramics, metals, and polymers when indented; see, e.g., the reviews by Lawn (1998), Guin and Gueguen (2019), and references therein. Yet, it was just very recently that the variational phase-field approach was used for the first time to study this kind of experimentally pervasive problems (Strobl and Seelig, 2020). In a nutshell, Strobl and Seelig (2020) first demonstrated that the variational phase-field model in its classical form fails to generate results that are consistent with experimental observations. As elaborated at length in Section 3 of Kumar et al. (2020), the reason for this drawback lies in the fact that the variational phase-field approach to fracture does not account for one of the required three basic ingredients to model fracture nucleation: \emph{the strength surface of the material}. They then proposed a modified phase-field model where \emph{de facto} only certain ``tensile'' part $W^+(\bfE)$ of the elastic energy $W(\bfE)$ is involved in the fracture process and where the localization length $\varepsilon$ is imparted physical meaning by tying up its value to the uniaxial tensile strength of the material at hand. This modified approach led to much improved predictions, although not without deficiencies, chief among these being the inability of the model to deal with arbitrary values of uniaxial tensile strength. As also elaborated at length in Section 3 of Kumar et al. (2020), the reason for this shortcoming lies in the fact that modifications of the variational phase-field approach to fracture based on imparting physical meaning to the localization length $\varepsilon$ are incomplete because their purely energetic character render them incapable of describing strength surfaces at large. By construction, the revisited phase-field formulation (\ref{BVP-u-theory})--(\ref{BVP-v-theory}) is free of such shortcomings.

\subsection{Calibration of the material inputs entering the theory}

\paragraph{Elasticity.} Based on their own ultrasound measurements, Mouginot and Maugis (1985) determined the Young's modulus and Poisson's ratio of their borosilicate glass to be $E=80$ GPa and $\nu=0.22$. We hence set the Lam\'e constants to
\begin{equation*}
\lambda=\dfrac{E\nu}{(1+\nu)(1-2\nu)}=26\, {\rm GPa}
\end{equation*}
and
\begin{equation}
\mu=\dfrac{E}{2(1+\nu)}=33\, {\rm GPa}\label{W-glass}
\end{equation}
in our simulations.

\begin{figure}[t!]
		\centering
		\includegraphics[width=0.90\linewidth]{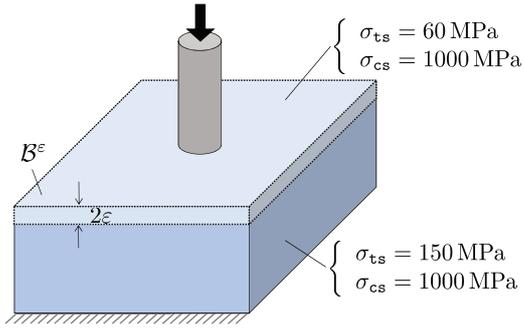}\vspace{-0.1cm}
		\caption{Schematic of the boundary layer $\mathcal{B}^\varepsilon$ wherein the glass strength is taken to be smaller than that in the bulk because of the presence of larger defects.}\label{Fig2}
\end{figure}

\paragraph{Strength.} No direct experimental data on the borosilicate glass strength was provided by Mouginot and Maugis (1985). However, they did estimate that the roughness of the surfaces of the specimens, due to their abrading treatment, was of a few microns. This size of boundary defects is much larger than the typical nanometer size of the inherent defects in the bulk of glass. The strength of the glass on the boundary of the specimens is thus expected to be significantly smaller than that of the glass in the bulk of the same specimens. In our simulations, accordingly, we take the strength of the glass to be characterized by the piecewise-constant Drucker-Prager strength surface
\begin{equation}
\mathcal{F}=\sqrt{J_2}+\dfrac{\sigma_{\texttt{cs}}-\sigma_{\texttt{ts}}}{\sqrt{3}(\sigma_{\texttt{cs}}+\sigma_{\texttt{ts}})} I_1-
\dfrac{2\scs\sts}{\sqrt{3}(\sigma_{\texttt{cs}}+\sigma_{\texttt{ts}})}=0\label{F-DP-Glass}
\end{equation}
with
\begin{equation}
\left\{\begin{array}{ll} \left\{\begin{array}{l}\sigma_{\texttt{ts}}=60\, {\rm MPa}\\ \sigma_{\texttt{cs}}=1000\, {\rm MPa}\end{array}\right., & \bfX\in \mathcal{B}^\varepsilon \\ \\ \left\{\begin{array}{l}\sigma_{\texttt{ts}}=150\, {\rm MPa} \\ \sigma_{\texttt{cs}}=1000\, {\rm MPa}\end{array}\right., & \bfX \in {\mathrm \Omega}\setminus\mathcal{B}^\varepsilon
\end{array}\right. ,\label{sts-scs-Glass}
\end{equation}
where $\mathcal{B}^\varepsilon$ is the $2\varepsilon$-thick boundary layer depicted in Fig. \ref{Fig2}, and where the values of the uniaxial tensile $\sigma_{\texttt{ts}}$ and compressive $\sigma_{\texttt{cs}}$ strengths are estimates from Fig. 7.29 in the review chapter by Guin and Gueguen (2019) on mechanical properties of glass.

\begin{remark}{\rm The choice of $2\varepsilon$ as the thickness of the boundary layer $\mathcal{B}^\varepsilon$ is dictated by the fact that smaller thicknesses would go undetected by the $\varepsilon$-regularized theory (\ref{BVP-u-theory})--(\ref{BVP-v-theory}). On the other hand, numerical experiments show that larger thicknesses do not significantly alter when or how cracks nucleate.
}
\end{remark}

\begin{remark}{\rm For definiteness, we do not consider here the full stochasticity of the strength of the glass. Within the proposed theory, however, it is a simple matter to do so by allowing the uniaxial tensile $\sigma_{\texttt{ts}}$ and compressive $\sigma_{\texttt{cs}}$ strengths in the strength surface (\ref{F-DP-Glass}) to take on spatially random values about some averages throughout the specimens.
}
\end{remark}

\paragraph{Critical energy release rate.} Mouginot and Maugis (1985) did not carry out independent experiments to measure the critical energy release rate of their borosilicate glass. However, they did estimate it to be within the range $5\,{\rm N/m}\leq G_c \leq 10\,{\rm N/m}$ directly from their indentation experiments. In our simulations we use the value
\begin{equation}
G_c=9\, {\rm N/m}.\label{Gc-Glass}
\end{equation}

In our simulations, furthermore, we idealize the steel punch as rigid. Also, exploiting symmetry and the fact that the indenter radii $A=0.05, 0.10, 0.25$, $0.50, 1.00,$ and $2.50$ mm are much smaller than the dimensions $50 \,{\rm mm}\, \times 50 \,{\rm mm}\, \times 25.4 \,{\rm mm}$ of the glass specimens, we idealize the problem to be axisymmetric. Recalling from Section 4.3 in Kumar et al. (2020) that the actual size of the regularized cracks is given by the relation $\varepsilon^\star=\varepsilon/\sqrt{1+\delta^\varepsilon}$ and that the coefficient $\delta^\varepsilon$ depends on the strength material constants (\ref{sts-scs-Glass}), we set the localization length to the sufficiently small piecewise-constat value
\begin{equation*}
\varepsilon=\left\{\begin{array}{ll}
26\, \mu{\rm m}, &\; \bfX\in \mathcal{B}^\varepsilon \\ \\
7.5\, \mu{\rm m}, &\; \bfX \in {\mathrm \Omega}\setminus\mathcal{B}^\varepsilon
\end{array}\right. 
\end{equation*}
so that $\varepsilon^\star$ is one and the same in the entirety of the specimens. Indeed, for this localization length and the above-specified material parameters, the calibrated coefficient $\delta^\varepsilon=14.2$ in the boundary layer $\mathcal{B}^\varepsilon$ and $\delta^\varepsilon=2.62$ in the rest of the domain ${\mathrm \Omega}\setminus\mathcal{B}^\varepsilon$ so that $\varepsilon^\star\approx 6$ $\mu$m in the entirety of ${\mathrm \Omega}$. We carry out the simulations in an unstructured mesh of size $h=1.5$ $\mu$m $\approx\varepsilon^\star/4 $ around the indenter where the cracks are expected to nucleate and propagate.

\subsection{Theory vs. experiments}

We are now in a position to deploy the theory (\ref{BVP-u-theory})--(\ref{BVP-v-theory}) with external driving force (\ref{cehat}), specialized to the stored-energy function (\ref{Wiso}) with elastic material constants (\ref{W-glass}), Drucker-Prager strength surface (\ref{F-DP-Glass}) with piecewise-constant uniaxial tensile $\sigma_{\texttt{ts}}$ and compressive $\sigma_{\texttt{cs}}$ strengths (\ref{sts-scs-Glass}), and critical energy release rate (\ref{Gc-Glass}), to simulate the indentation experiments of Mouginot and Maugis (1985).

\begin{figure}[t!]
		\centering
		\includegraphics[width=0.7\linewidth]{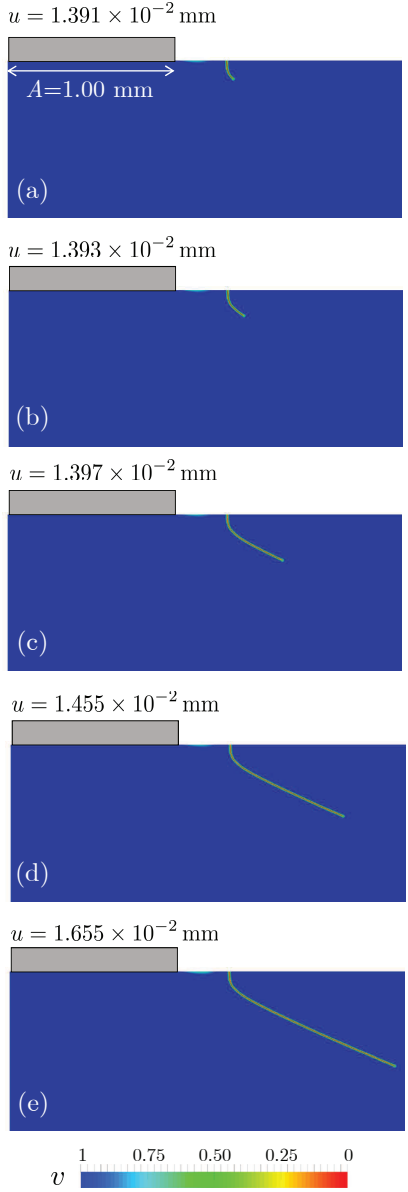}\vspace{-0.1cm}
		\caption{Contour plots of the phase field $v$ predicted by the theory in the specimen indented with the indenter of radius $A=1.00$ mm at five applied displacements $u$. Part (a) shows the instance at which fracture nucleates, while parts (b)--(e) show the ensuing propagation of the nucleated ring crack into a cone crack. }\label{Fig3}
\end{figure}
\begin{figure}[b!]
  \subfigure[]{
   \begin{minipage}[]{0.5\textwidth}
   \centering \includegraphics[width=0.8\linewidth]{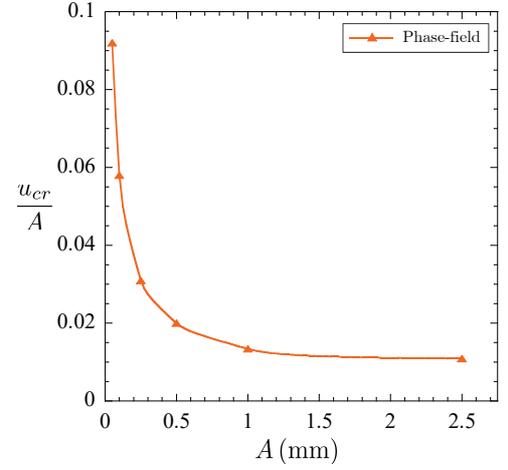}
   \vspace{0.2cm}
   \end{minipage}}
  \subfigure[]{
   \begin{minipage}[]{0.5\textwidth}
   \centering \includegraphics[width=0.8\linewidth]{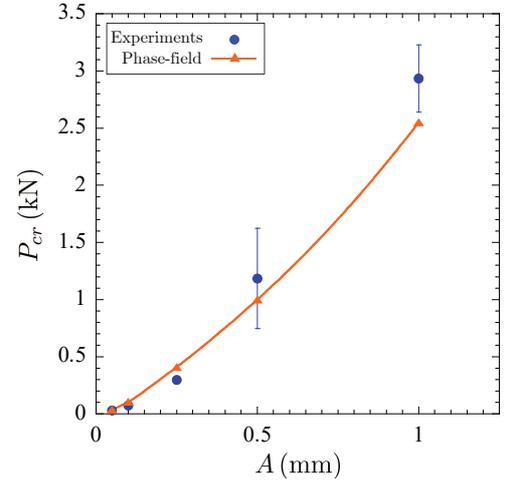}
   \vspace{0.2cm}
   \end{minipage}}
   \caption{Comparison between the predictions of the phase-field theory and the experiments of Mouginot and Maugis (1985) on borosilicate glass. (a) The normalized critical displacement $u_{cr}/A$ and (b) the critical force $P_{cr}$ at which the ring crack nucleates as functions of the radius $A$ of the indenter. Solid lines interpolating between the phase-field results are included to aid visualization.}\label{Fig4}
\end{figure}
Consistent with the experimental observations schematically depicted in Fig. \ref{Fig1}(b), irrespective of the indenter radius $A$, all the simulations exhibit the following two successive events:
\begin{itemize}

\item{fracture nucleation occurs roughly in the form of a ring crack of depth $H$ at a distance $R>A$ away from the centerline of the indenter at some critical value $u_{cr}$ of the applied displacement $u$,}

\item{subsequently, the crack proceeds its propagation, first rapidly and later on significantly more slowly, at a roughly constant angle $\theta$ with respect to the surface of the specimen, forming thus a cone crack.}

\end{itemize}
\begin{figure}[t!]
  \subfigure[]{
   \begin{minipage}[]{0.5\textwidth}
   \centering \includegraphics[width=0.8\linewidth]{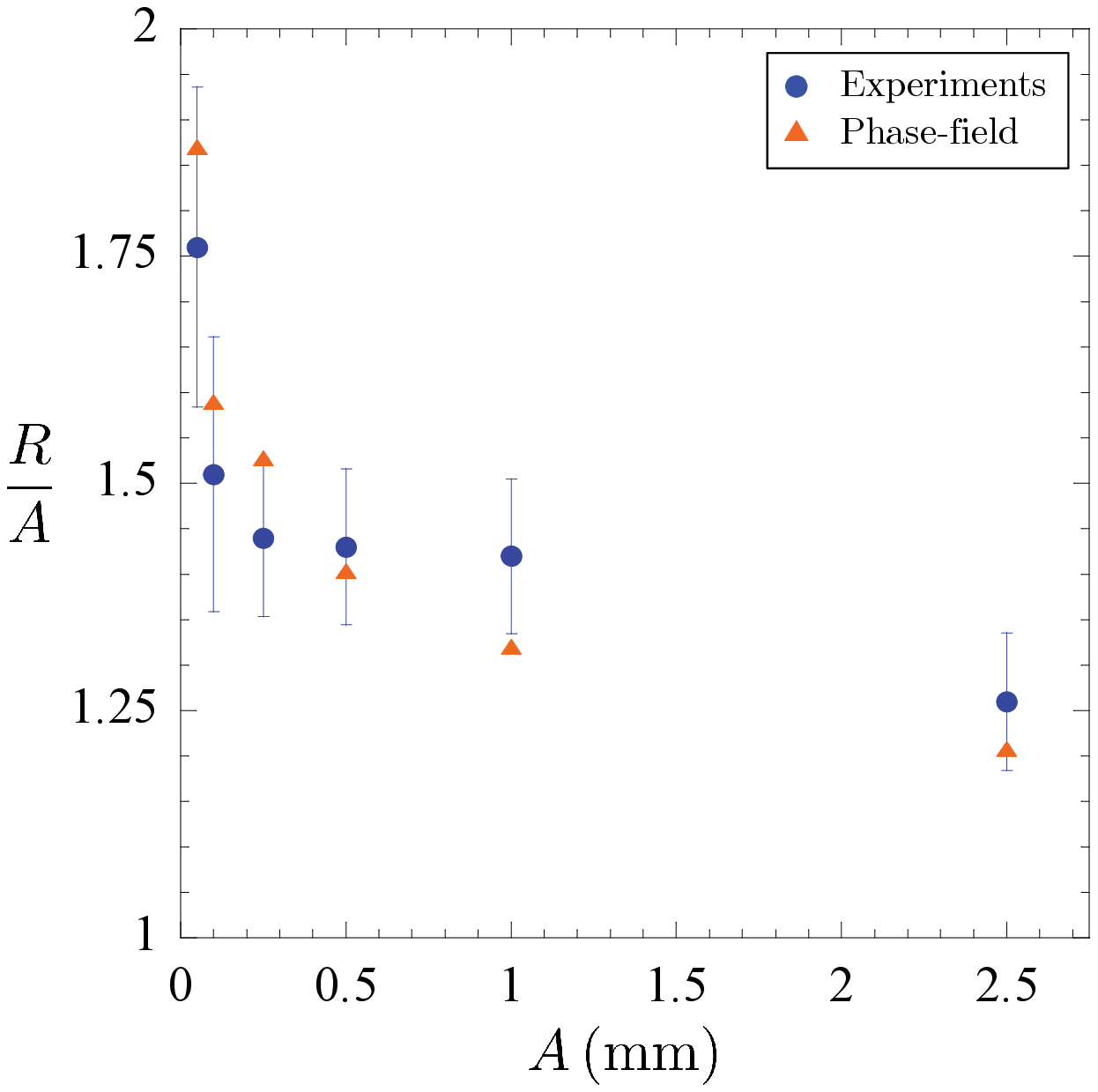}
   \vspace{0.2cm}
   \end{minipage}}
  \subfigure[]{
   \begin{minipage}[]{0.5\textwidth}
   \centering \includegraphics[width=0.8\linewidth]{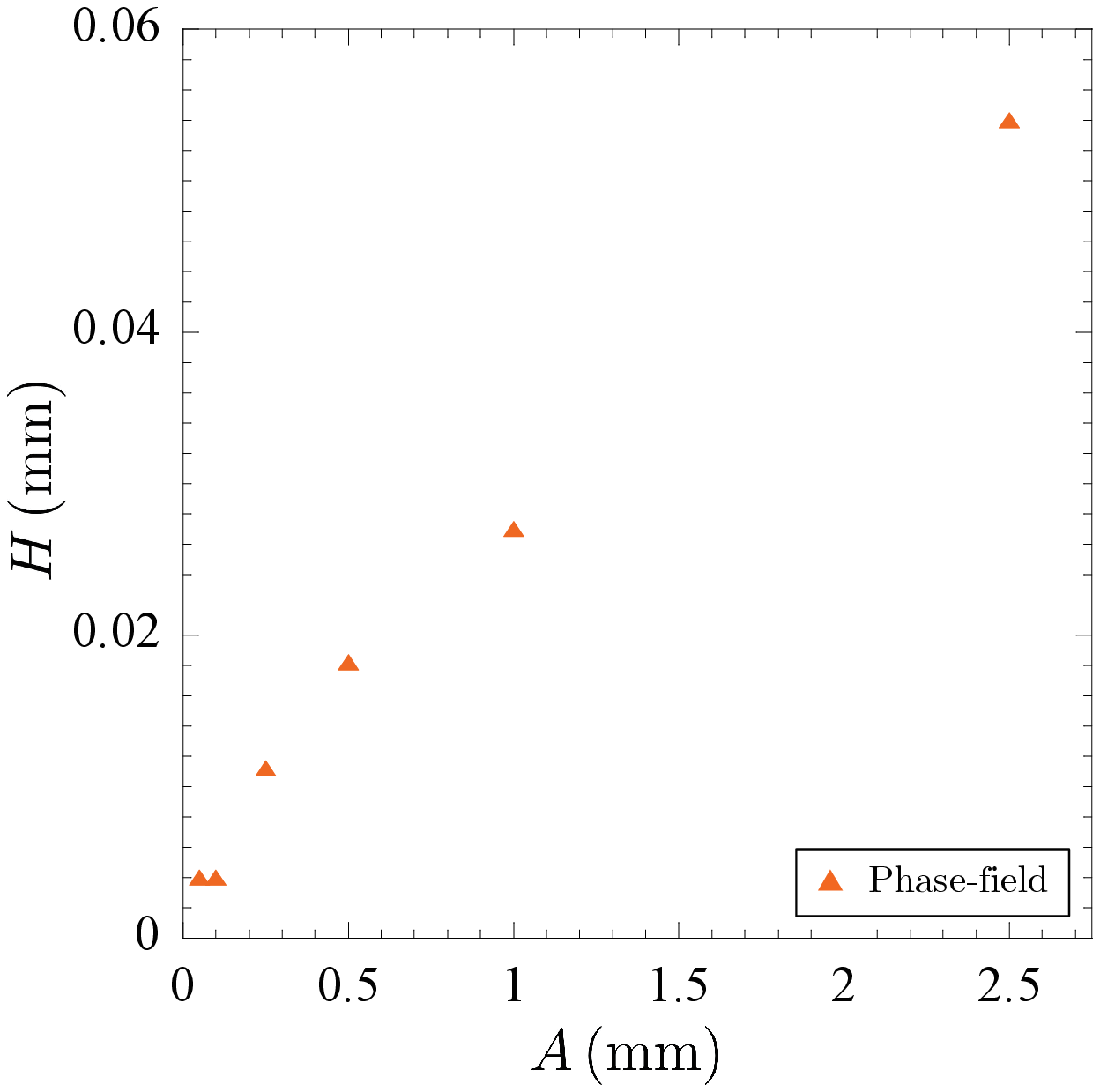}
   \vspace{0.2cm}
   \end{minipage}}
   \caption{Comparison between the predictions of the phase-field theory and the experiments of Mouginot and Maugis (1985) on borosilicate glass. (a) The normalized radial location $R/A$ at which the ring crack nucleates and (b) its depth $H$ as functions of the radius $A$ of the indenter.}\label{Fig5}
\end{figure}
\begin{figure}[t!]
  \subfigure[]{
   \begin{minipage}[]{0.5\textwidth}
   \centering \includegraphics[width=0.78\linewidth]{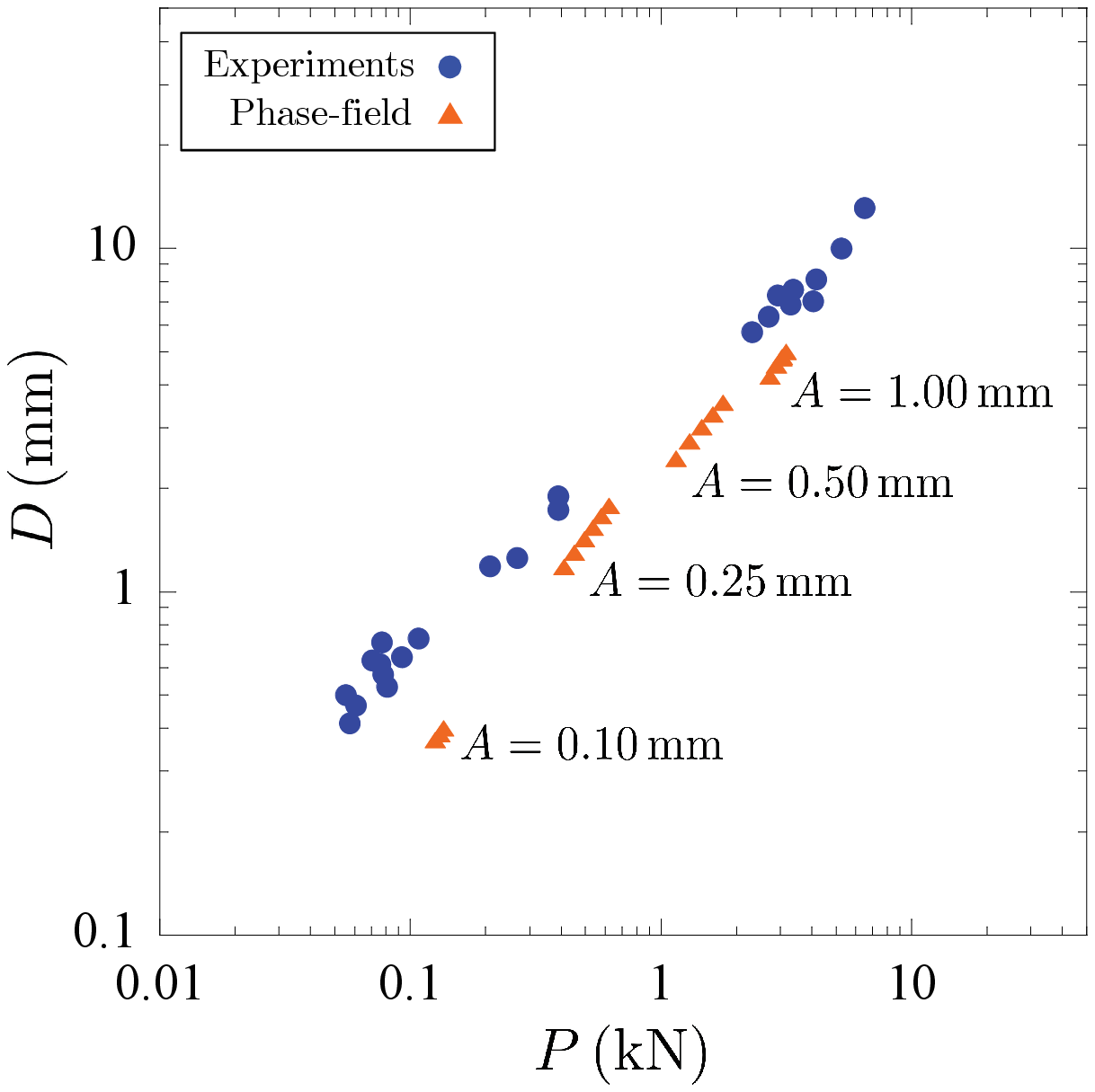}
   \vspace{0.2cm}
   \end{minipage}}
  \subfigure[]{
   \begin{minipage}[]{0.5\textwidth}
   \centering \includegraphics[width=0.78\linewidth]{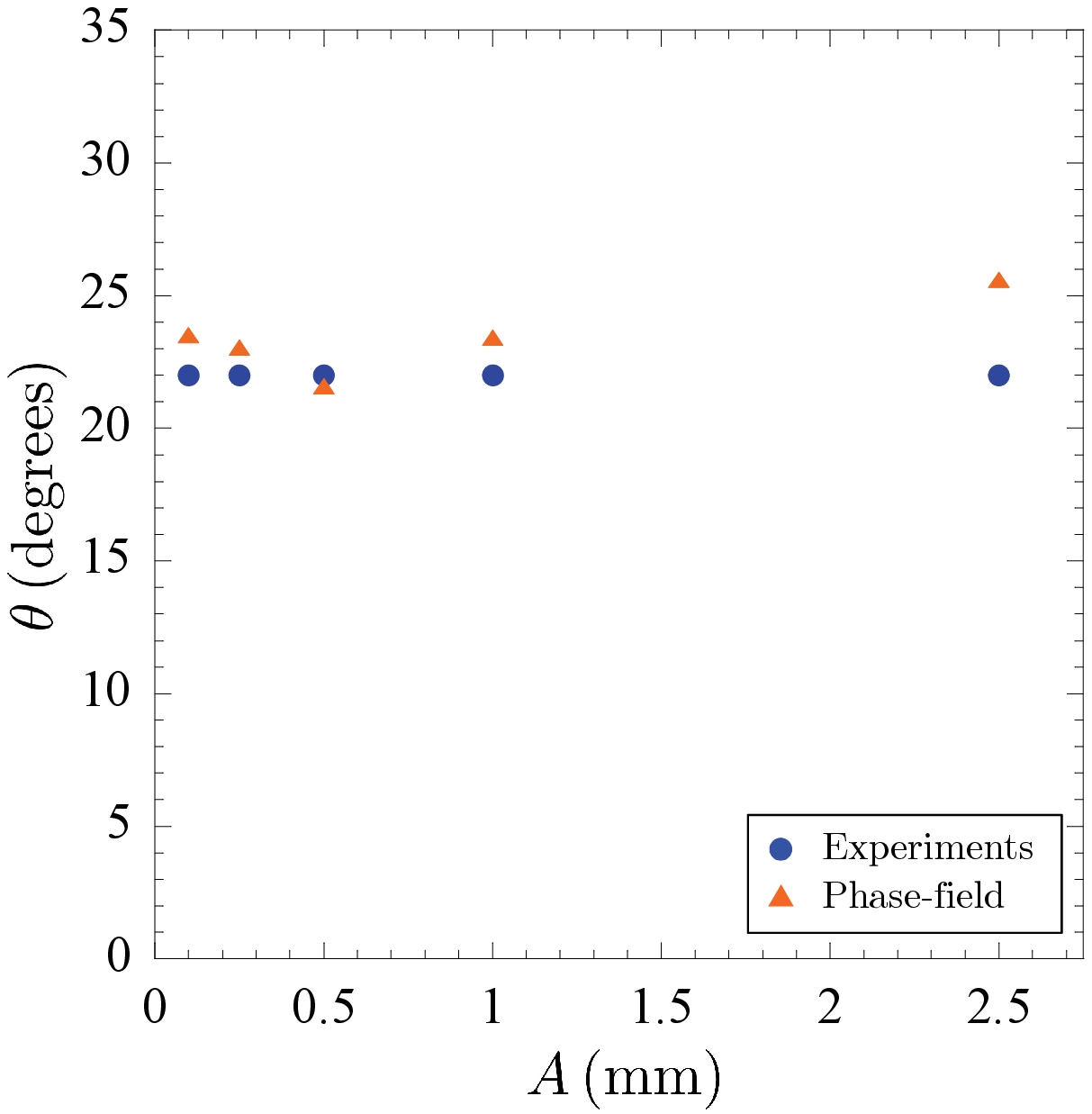}
   \vspace{0.2cm}
   \end{minipage}}
   \caption{Comparison between the predictions of the phase-field theory and the experiments of Mouginot and Maugis (1985) on borosilicate glass. (a) The growing diameter $D$ of the cone crack as a function of the force $P$ exerted by the indenter. (b) The angle $\theta$ of the cone crack as a function of the radius $A$ of the indenter; the experimental scatter in $\theta$ is not included because it was not reported by Mouginot and Maugis (1985).}\label{Fig6}
\end{figure}
Figure \ref{Fig3} shows representative snapshots of the phase field $v$ predicted by the theory for the case of the indenter with radius $A=1.00$ mm that illustrate the above-outlined two events. Here, it is important to remark that the nucleation of fracture occurs in a region where the stress field is not singular but sharply shifts from a simple shear stress state at the surface of the specimen to a fully triaxial stress state in its adjacent bulk; for contours of the stress field prior to fracture, see for instance Chapter 5 in the monograph by Fischer-Cripps (2007). Nucleation is thus controlled by the strength surface (\ref{F-DP-Glass}) in the boundary layer $\mathcal{B}^\varepsilon$ of the specimen, in particular, the neighborhood around the shear strength (where $\sigma_2=-\sigma_1$ and $\sigma_3=0$) in that surface, and \emph{not} by the uniaxial tensile strength $\sts$ of the material as often incorrectly suggested in the literature. Once nucleated, of course, the propagation of the crack is governed by the Griffith competition between the bulk elastic energy in the glass and its surface fracture energy.

In addition to the foregoing qualitative description of the fracture process, Mouginot and Maugis (1985) reported measurements of the radial locations $R$ where the cracks nucleated, the critical values $P_{cr}$ of the force $P$ exerted by the indenter at which they first observed a fracture event, the angle $\theta$ of the eventual cone cracks, as well as of their growing diameter $D$ as a function of the indenter force $P$. Figures \ref{Fig4}, \ref{Fig5}, and \ref{Fig6}  present comparisons between the predictions generated by the phase-field theory and such measurements. Since Mouginot and Maugis (1985) did not include results for the critical displacement $u_{cr}$ or the depth $H$ of the ring cracks, Figs. \ref{Fig4}(a) and \ref{Fig5}(b) only present the predictions from the theory. Also, since Mouginot and Maugis (1985) did not make precise what was the fracture event that they first observed, for definiteness, the theoretical predictions for $P_{cr}$ that are plotted in Fig. \ref{Fig4}(b) correspond to the values of the indenter force $P$ at which the ring crack nucleates; it is likely that this event was indeed their first observation since the crack is several microns in length by then.

It is plain from these comparisons that the theory is in fairly good quantitative agreement with all the experimental results. It should be noted, however, that the critical force $P_{cr}=0.03$ kN measured from the experiments with the smallest indenter of radius $A=0.05$ mm is likely too large for the steel punch not to have yielded (since $P_{cr}/\pi A^2=3.82$ GPa) and deformed substantially, as opposed to have remained undeformed as assumed in the simulations. It is also noteworthy that the experimental data features quite a bit of scatter, especially for the quantities $R/A$ and $P_{cr}$ associated with the nucleation. Presumably, this is because of the stochasticity introduced by the abrasion treatment of the surface of the specimens. Again, as noted above, such a stochasticity can be incorporated in the theory by having the strength material constants (here, $\sigma_{\texttt{ts}}$ and $\sigma_{\texttt{cs}}$) in the strength surface (\ref{F-DP-Glass}) to take on spatially random values about some average. We do not include examples of such simulations here but, instead, refer the interested reader to Section 5 in Kumar and Lopez-Pamies (2020) for examples in the context of nucleation of fracture in silicone elastomers where strength stochasticity plays a key role.

\section{Application to the three-point bending of U-notched PMMA beams} \label{Sec: U-Notch}

\begin{figure}[h!]
		\centering
		\includegraphics[width=0.85\linewidth]{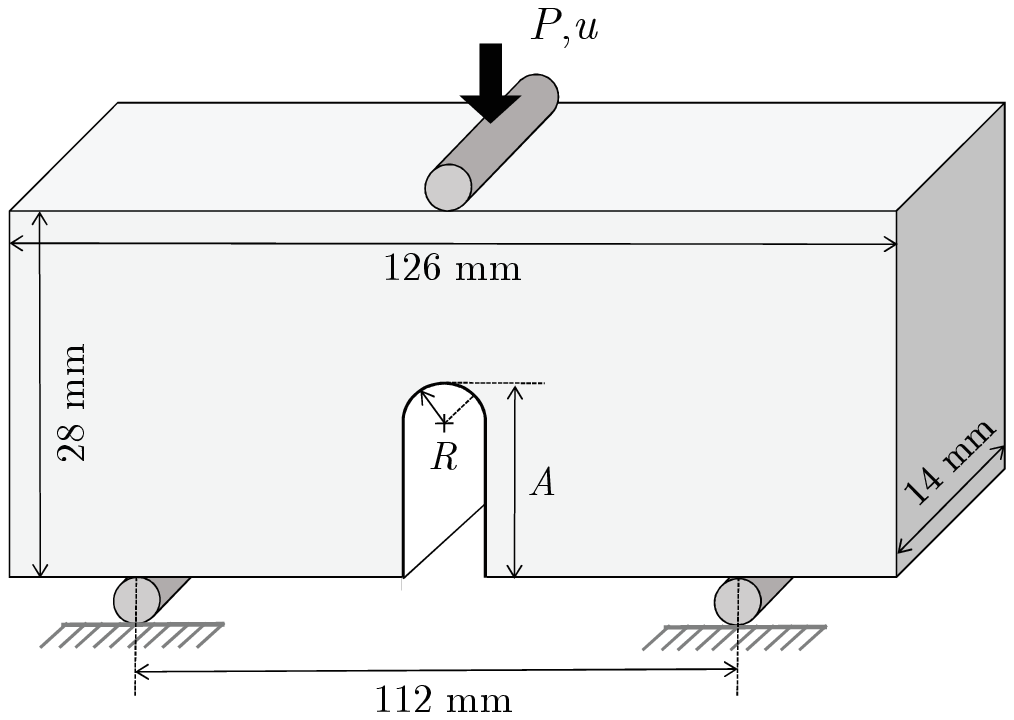}\vspace{-0.1cm}
		\caption{Schematic of the initial specimen geometry and applied boundary conditions for the three-point bending experiments carried out by G\'omez et al. (2005) on U-notched PMMA beams at $-60$ $^{\circ}$C.}\label{Fig7}
\end{figure}

Next, we confront the theory to the experiments of G\'omez et al. (2005) for the three-point bending of U-notched PMMA beams. Figure \ref{Fig7} shows a schematic of the initial geometry of the specimens and the applied boundary conditions. Results were reported for specimens with U notches of two depths, approximately $A=5$ and $14$ mm, and seven radii, approximately $R=0.18$, $0.34$, $0.52$, $0.94$, $1.47$, $1.97$, and $3.98$ mm. The specimens were subjected to a prescribed displacement $u$, which was applied at the fixed rate of $\dot{u}=0.5$ $\mu$m/s; the corresponding force is denoted by $P$. With the objective of minimizing deviations from linear elastic brittle behavior, all the experiments were carried out at the low temperature of $-60$ $^{\circ}$C.

\subsection{Calibration of the material inputs entering the theory}

\paragraph{Elasticity.} Based on their own uniaxial tension tests on carefully machined cylindrical samples, G\'omez et al. (2005) determined the Young's modulus and Poisson's ratio of their PMMA at $-60$ $^{\circ}$C to be roughly $E=5.05$ GPa and $\nu=0.40$. Therefore, the Lam\'e constants in our simulations are set to
\begin{equation*}
\lambda=\dfrac{E\nu}{(1+\nu)(1-2\nu)}=7.21\, {\rm GPa}
\end{equation*}
and
\begin{equation}
\mu=\dfrac{E}{2(1+\nu)}=1.80\, {\rm GPa}.\label{W-PMMA60}
\end{equation}

\paragraph{Strength.} From the same uniaxial tension tests used to determine the elastic constants, G\'omez et al. (2005) also estimated the uniaxial tensile strength to be roughly $\sts=128\, {\rm MPa}$. In our simulations, accordingly, we take the strength of the PMMA at $-60$ $^{\circ}$C to be characterized by the Drucker-Prager strength surface (\ref{F-DP-Glass}) with
\begin{equation}
\left\{\begin{array}{l}\sigma_{\texttt{ts}}=128\, {\rm MPa}\\[5pt] \sigma_{\texttt{cs}}=256\, {\rm MPa}\end{array}\right. ,\label{sts-scs-PMMA60}
\end{equation}
where the compressive strength $\sigma_{\texttt{cs}}$ is set to be twice as large as the tensile strength $\sigma_{\texttt{ts}}$ in accordance with typical values at room temperature found elsewhere in the literature.

\begin{remark}{\rm G\'omez et al. (2005) stated that the U notches were carefully machined in their specimens. We therefore assume here that the sizes of the inherent defects in the bulk and on the boundary of the specimens are not fundamentally different and hence that the strength surface (\ref{F-DP-Glass}) with (\ref{sts-scs-PMMA60}) applies to the entirety of the specimens. Much like in the preceding comparisons with glass, we also adopt in our simulations the idealization that the strength material constants (\ref{sts-scs-PMMA60}) are deterministic as opposed to stochastic.
}
\end{remark}

\paragraph{Critical energy release rate.}  From their own compact and single-edge-notch tension tests, G\'omez et al. (2005) measured the critical energy release rate of their PMMA at $-60$ $^{\circ}$C to be roughly
\begin{equation}
G_c=480\, {\rm N/m}.\label{Gc-PMMA60}
\end{equation}
We use this value in our simulations.

We make use of the sufficiently small localization length $\varepsilon=25$ $\mu$m in all our simulations. For such a localization length and the above-specified material parameters, the calibrated coefficient $\delta^\varepsilon$ in the external driving force (\ref{cehat}) takes the value $\delta^\varepsilon=2.15$. The size of the regularized cracks is hence given by $\varepsilon^\star=\varepsilon/\sqrt{1+\delta^\varepsilon}\approx 14$ $\mu$m. We carry out the simulations under plane-strain conditions in an unstructured mesh of size $h=3$ $\mu$m $\approx\varepsilon^\star/5$ around and ahead of the U-notch, where the cracks are expected to nucleate and propagate. Although the elastic fields are fully triaxial (neither plane-strain nor plane-stress), the plane-strain idealization is justified by 3D simulations that show that the results do not differ significantly from those of plane strain.

\subsection{Theory vs. experiments}
\begin{figure}[b!]
		\centering
		\includegraphics[width=0.85\linewidth]{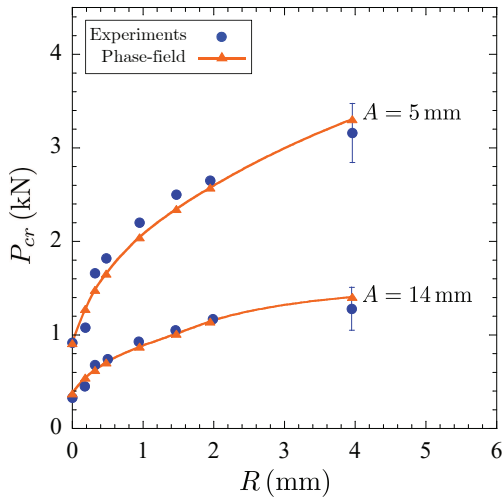}\vspace{-0.1cm}
		\caption{Comparison between the predictions of the phase-field theory and the experiments of G\'omez et al. (2005) on PMMA at $-60$ $^{\circ}$C for the critical force $P_{cr}$ at which fracture nucleates as a function of the notch radius $R$. Results are shown for the two notch depths $A=5$ and $14$ mm.}\label{Fig8}
\end{figure}
All the specimens tested by G\'omez et al. (2005) showed a linear force-displacement ($P$ versus $u$) response followed by complete rupture. Figure \ref{Fig8} compares the predictions generated by the phase-field theory with their experimental data for the maximum force $P_{cr}$ reached in those force-displacement responses, which presumably indicate the points at which fracture nucleated. The results are shown as a function of the notch radius $R$ for the two notch depths $A=5$ and $14$ mm that they considered.

\begin{figure}[b!]
		\centering
		\includegraphics[width=0.65\linewidth]{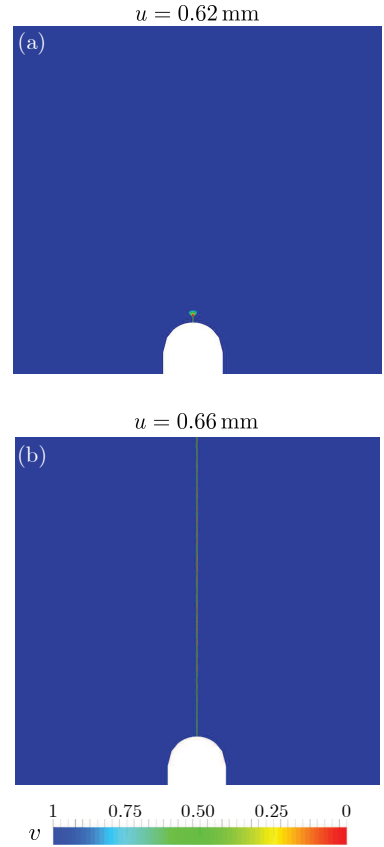}\vspace{-0.1cm}
		\caption{Contour plots of the phase field $v$ predicted by the theory in the U-notched beam with notch radius $R=0.94$ mm and notch depth $A=5$ mm at two applied displacements $u$. Part (a) shows the instance at which fracture nucleates on the tip of the notch, while part (b) shows the instance at which the specimen is all but severed into two pieces.}\label{Fig9}
\end{figure}
Two comments are in order. First, the theory is in good quantitative agreement with the experimental results. Second, as expected, the results distinctly show that fracture nucleation transitions from being governed by the Griffith competition between bulk elastic energy and surface fracture energy to being governed by the strength of the material as the notch radius $R$ increases.

Unfortunately, G\'omez et al. (2005) did not include results on the location of fracture nucleation or details of its subsequent propagation. According to the theory, irrespective of the notch radius and depth, fracture nucleation occurs at the tip of the notch and propagates rapidly (more so the larger the notch radius) along the center plane of the specimens severing them into two halves. For completeness, Fig. \ref{Fig9} shows representative snapshots of the phase field $v$ for the specimen with notch radius $R=0.94$ mm and notch depth $A=5$ mm at two values of the applied displacement $u$. Figure \ref{Fig9}(a) shows the instance right after fracture nucleation, while Fig. \ref{Fig9}(b) shows the instance right before the specimen is severed into two pieces.

\begin{remark}{\rm Here, it is important to note that the nucleation of fracture in U-notched beams subject to three-point bending need not necessarily occur at the notch tip. For instance, for materials with weak hydrostatic strength, fracture may nucleate within the bulk ahead of the notch tip; see, e.g., the experiments of Aranda-Ruiz et al. (2020) on polycarbonate. According to the Drucker-Prager strength surface (\ref{F-DP-Glass}) with (\ref{sts-scs-PMMA60}) assumed here for PMMA at $-60$ $^{\circ}$C --- for which the hydrostatic strength $\sigma_{\texttt{hs}}=2\scs\sts/3(\scs-\sts)=171$ MPa is significantly larger than the uniaxial tensile strength $\sts=128$ MPa --- the nucleation of fracture always occurs at the notch tip.
}
\end{remark}

\section{Application to the three-point bending of V-notched PMMA beams} \label{Sec: V-Notch}

\begin{figure}[b!]
		\centering
		\includegraphics[width=0.95\linewidth]{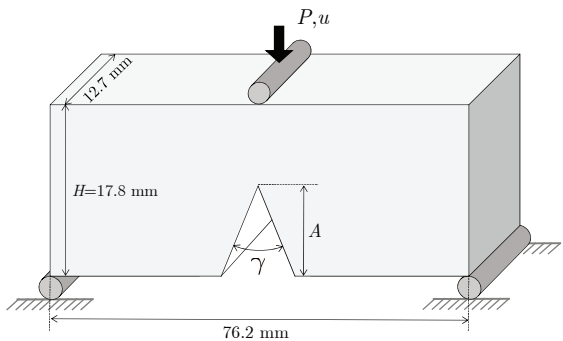}\vspace{-0.1cm}
		\caption{Schematic of the initial specimen geometry and applied boundary conditions for the three-point bending experiments carried out by Dunn et al. (1997) on V-notched PMMA beams.}\label{Fig10}
\end{figure}

Finally, we turn to confronting the theory to the experiments of Dunn et al. (1997) for the three-point bending of V-notched PMMA beams. A schematic of the initial geometry of the specimens and the applied boundary conditions used by these authors is shown in Fig. \ref{Fig10}. They reported results for specimens with V notches of four depths, $A=1.78$, $3.56$, $5.33$, and $7.11$ mm, and three angles, $\gamma=60^{\circ}$, $90^{\circ}$, and $120^{\circ}$. The specimens were subjected to a prescribed displacement $u$ that was applied at the fixed rate of $\dot{u}=100$ $\mu$m/s; the corresponding force is denoted by $P$. In contrast to the experiments of G\'omez et al. (2005) discussed in the preceding section, Dunn et al. (1997) performed all their experiments at room temperature.

\subsection{Calibration of the material inputs entering the theory}

\paragraph{Elasticity.} From their own uniaxial tension tests, Dunn et al. (1997) determined the Young's modulus of their PMMA to be $E=2.3$ GPa. While they did not measure the Poisson's ratio, they referred to a private communication from Ledbetter (1996) to assert that $\nu=0.36$. Accordingly, in our simulations, we make use of the Lam\'e constants
\begin{equation*}
\lambda=\dfrac{E\nu}{(1+\nu)(1-2\nu)}=2.17\, {\rm GPa}
\end{equation*}
and
\begin{equation}
\mu=\dfrac{E}{2(1+\nu)}=0.85\, {\rm GPa}\label{W-PMMA}
\end{equation}

\paragraph{Strength.} In order to measure the uniaxial tensile strength of their PMMA, Dunn et al. (1997) also carried out three-point bending of beams without notches. Those indicated that $\sts=124\pm 20$ MPa. They did not report any other strength data.  In our simulations, we take the strength of their PMMA to be characterized by the Drucker-Prager strength surface (\ref{F-DP-Glass}) with
\begin{equation}
\left\{\begin{array}{l}\sigma_{\texttt{ts}}=124\, {\rm MPa}\\[5pt] \sigma_{\texttt{cs}}=248\, {\rm MPa}\end{array}\right.,\label{sts-scs-PMMA}
\end{equation}
where, for the same reasons invoked in the preceding section, we have set the uniaxial compressive strength $\sigma_{\texttt{cs}}$ to be twice as large as the tensile strength $\sigma_{\texttt{ts}}$.

\begin{remark}{\rm Dunn et al. (1997) mentioned that the machining of the notches was done carefully and that optical microscopy revealed no evidence of crazing
ahead of the machined notch tips. In our simulations, we therefore assume that the strength surface (\ref{F-DP-Glass}) with (\ref{sts-scs-PMMA}) applies to the entirety of the specimens, the bulk as well as the boundaries, including the (macroscopically) sharp notch tip. Furthermore, exactly as in the two sets of preceding comparisons, we also adopt here the idealization that the strength material constants (\ref{sts-scs-PMMA}) are deterministic and not stochastic.
}
\end{remark}

\paragraph{Critical energy release rate.} Using cracked three-point bending specimens with the same dimensions depicted in Fig. \ref{Fig10}, Dunn et al. (1997) also measured the critical energy release rate of their PMMA to be $G_c=394\pm 90\, {\rm N/m}$. In our simulations we use the value
\begin{equation}
G_c=394\, {\rm N/m}.\label{Gc-PMMA}
\end{equation}

As for the remainder of parameters entering the simulations, we make use of the localization length $\varepsilon=12.5$ $\mu$m, which is sufficiently small for this problem. For such a localization length and the above-calibrated material parameters, the coefficient $\delta^\varepsilon$ in the external driving force takes the value $\delta^\varepsilon=0.47$. The size of the regularized cracks is hence given by $\varepsilon^\star=\varepsilon/\sqrt{1+\delta^\varepsilon}\approx 10.3$ $\mu$m. We carry out the simulations under plane-strain conditions in an unstructured mesh of size $h=2$ $\mu$m $\approx\varepsilon^\star/5$ around and ahead of the V-notch, where the cracks are expected to nucleate and propagate. The plane-strain idealization, as it was the case for the U-notched specimens considered above, was checked to be sufficiently accurate via direct comparisons with 3D simulations.

\subsection{Theory vs. experiments}

\begin{figure}[t!]
		\centering
		\includegraphics[width=0.65\linewidth]{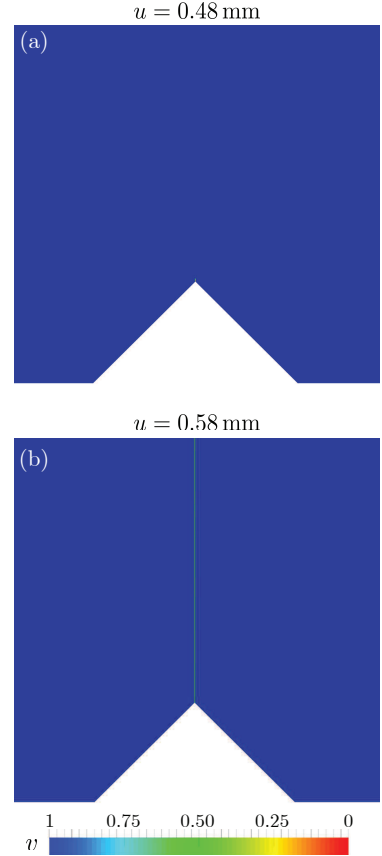}\vspace{-0.1cm}
		\caption{Contour plots of the phase field $v$ predicted by the theory in the V-notched beam with notch depth $A=3.56$ mm and notch angle $\gamma=90^{\circ}$ at two applied displacements $u$. Part (a) shows the instance at which fracture nucleates on the tip of the notch, while part (b) shows the instance at which the specimen is all but severed into two pieces.}\label{Fig11}
\end{figure}

Dunn et al. (1997) indicated that all their specimens showed a linear force-displacement ($P$ versus $u$) response followed by complete rupture into two symmetric pieces. Upon inspection of the mirror-like smooth fractured surfaces of the post-mortem specimens, they concluded that fracture nucleated at the notch tip and rapidly propagated along the center plane of the specimens, this irrespective of the depth and angle of the notch. All the simulations are in accordance with these observations. For illustration purposes, Figs. \ref{Fig11}(a) and \ref{Fig11}(b) show representative snapshots of the phase field $v$ predicted by the theory for the case of the V-notched beam with notch depth $A=3.56$ mm and notch angle $\gamma=90^{\circ}$ just after nucleation and just before complete rupture.

\begin{figure}[t!]
  \subfigure[]{
   \begin{minipage}[]{0.5\textwidth}
   \centering \includegraphics[width=0.73\linewidth]{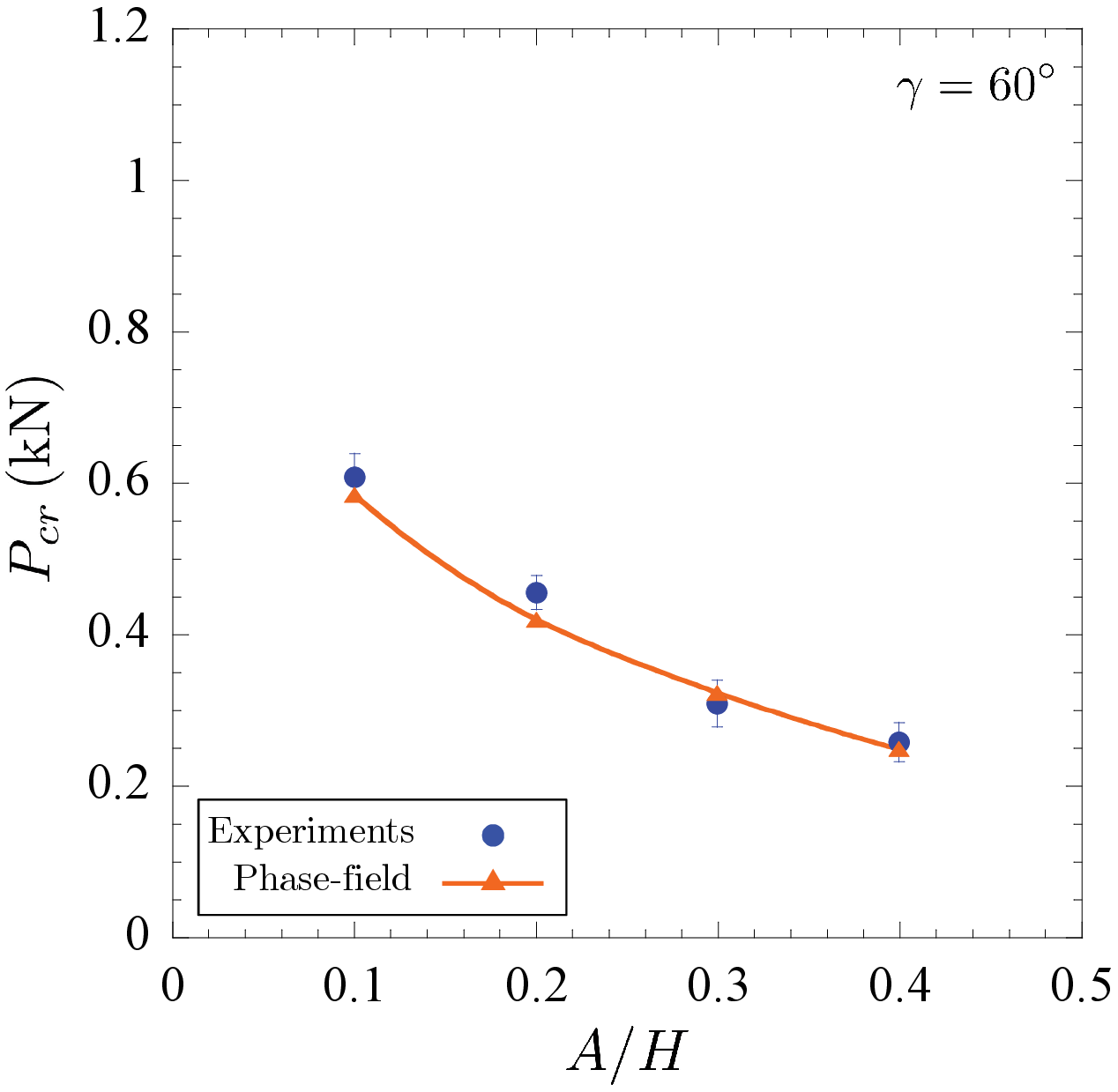}
   \vspace{0.2cm}
   \end{minipage}}
  \subfigure[]{
   \begin{minipage}[]{0.5\textwidth}
   \centering \includegraphics[width=0.73\linewidth]{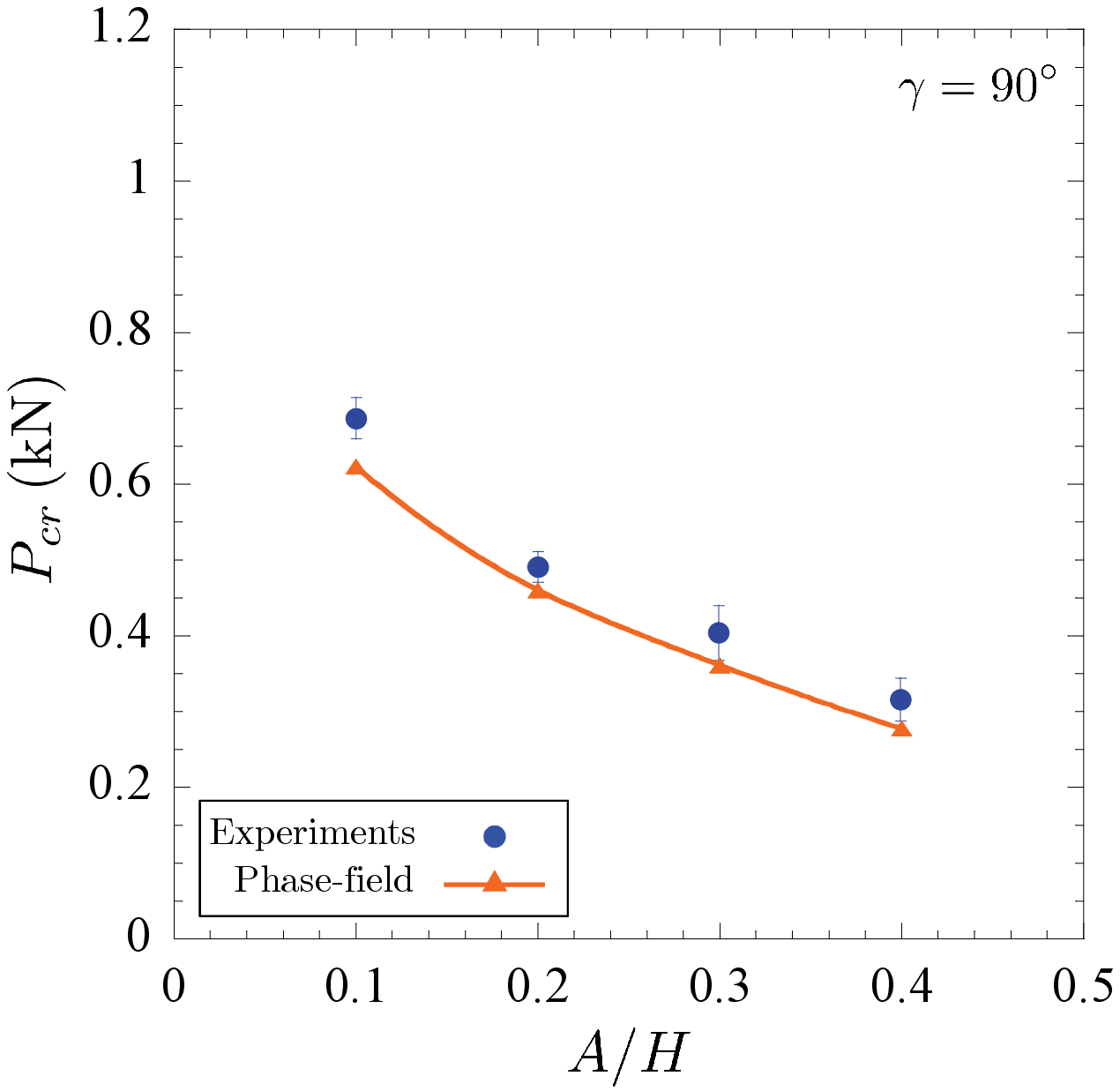}
   \vspace{0.2cm}
   \end{minipage}}
     \subfigure[]{
   \begin{minipage}[]{0.5\textwidth}
   \centering \includegraphics[width=0.73\linewidth]{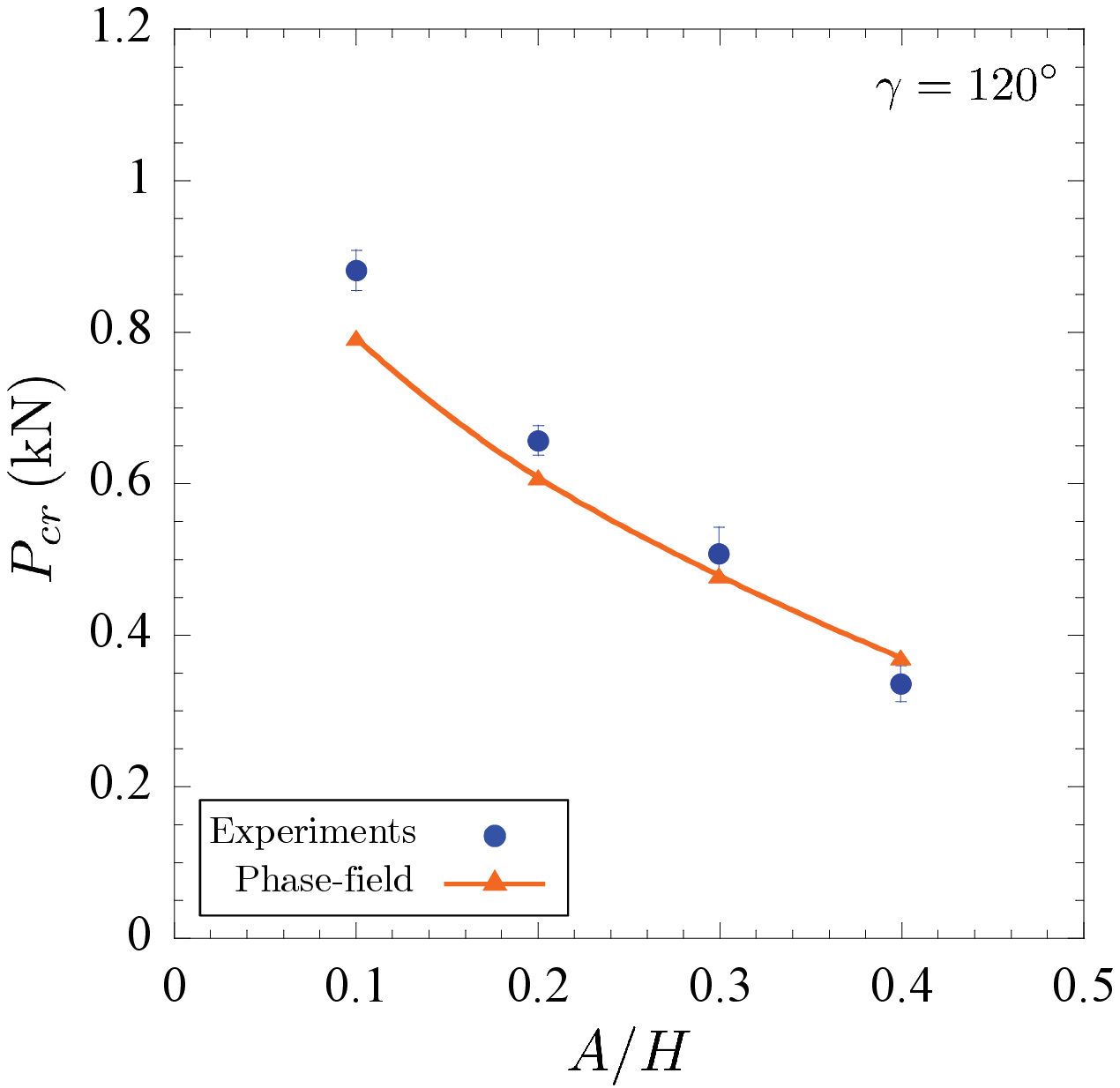}
   \vspace{0.2cm}
   \end{minipage}}
   \caption{Comparison between the predictions of the phase-field theory and the experiments of Dunn et al. (1997) on PMMA. The critical forces $P_{cr}$ at which fracture nucleates in specimens with notch angles (a) $\gamma=60^{\circ}$, (b) $\gamma=90^{\circ}$, and (c) $\gamma=120^{\circ}$, as functions of the normalized notch depth $A/H$.}\label{Fig12}
\end{figure}

Figure \ref{Fig12} presents comparisons between the predictions generated by the theory and the measurements of Dunn et al. (1997) for the critical forces $P_{cr}$ at which fracture nucleated. The results are shown as a function of the notch depth $A$, normalized by the height $H=17.8$ mm of the specimens, for the three notch angles $\gamma=60^{\circ}$, $90^{\circ}$, and $120^{\circ}$ that they investigated. In addition to the good agreement between the theory and the experiments, these comparisons show, as expected, that fracture nucleation transitions from being governed by the Griffith competition between bulk elastic energy and surface fracture energy to being governed by the strength of the material as the notch angle $\gamma$ increases.

\section{Final comments} \label{Sec: Final Comments}

Adding to the validation results presented in the companion papers Kumar et al. (2018a,b; 2020) and Kumar and Lopez-Pamies (2021), where the theory is confronted to experiments of fracture nucleation and propagation in silicone elastomers, titania, graphite, polyurethane, PMMA, alumina, and natural rubber spanning a broad spectrum of specimen geometries and loading conditions, the three sets of comparisons with experiments on glass and PMMA presented in this work provide further evidence and motivation to continue investigating the phase-field theory (\ref{BVP-u-theory})--(\ref{BVP-v-theory}) as a complete framework for the description of fracture nucleation and propagation in elastic brittle materials at large.

The results presented in this work have also made it plain that more experiments are pressingly needed to measure the strength surface of materials beyond the two standard points of uniaxial tensile and compressive strength. Indeed, fracture nucleation occurs more often than not in regions where the state of stress is fully triaxial. Any hope of being able to accurately predict fracture nucleation under general loading conditions appears then to hinge on having a more complete experimental knowledge of the strength surface of the materials at hand.

\section*{Acknowledgements}

\noindent Support for this work by the National Science Foundation through the grant CMMI--2132528 and the collaborative Grants CMMI--1901583 and CMMI--1900191 is gratefully acknowledged.

\end{document}